\documentclass[prb,superscriptaddress,twocolumn,showpacs,preprintnumbers,amsmath,amssymb,floatfix]{revtex4}

\usepackage{graphicx}
\usepackage{dcolumn}
\usepackage{bm}

\begin{document}

\preprint{APS/123-QED}

\title{
Magnetic field-induced one-magnon Raman scattering 
in the magnon Bose-Einstein condensation phase of TlCuCl$_{3}$
}

\author{Haruhiko Kuroe}
 \email{kuroe@sophia.ac.jp}
\author{Kouhei Kusakabe}%
\author{Akira Oosawa}%
\author{Tomoyuki Sekine}%
\affiliation{%
Department of Physics, Sophia University, %
7-1 Kioi-cho, Chiyoda-ku, Tokyo 102-8554, Japan}

\author{Fumiko Yamada}
\author{Hidekazu Tanaka}
\affiliation{%
Department of Physics, Tokyo Institute of Technology, %
Oh-okayama, Meguro-ku, Tokyo 152-8551, Japan}

\author{Masashige Matsumoto}
\affiliation{%
Department of Physics, Sizuoka University, %
836 Oya, Shizuoka 422-8529, Japan}

\date{\today}

\begin{abstract}
We report the observation of the $A_{\rm g}$-symmetric 
one-magnon Raman peak in the magnon Bose-Einstein condensation phase of TlCuCl$_{3}$.
Its Raman shift traces the one-magnon energy at the magnetic $\Gamma$ point, 
and its intensity is proportional to the squared transverse magnetization.
The appearance of the one-magnon Raman scattering 
originates from the exchange magnon Raman process
and reflects the change of the magnetic-state symmetry.
Using the bond-operator representation, 
we theoretically clarify the Raman selection rules, 
being consistent with the experimental results.

\end{abstract}

\pacs{78.30.-j,75.10.Jm}
\maketitle

\section{\label{Introduction}Introduction}

Currently, many physicists are examining 
the Bose-Einstein condensation (BEC) of atoms in ultracooled dilute gases, 
and in particular, the BEC of magnons.
The latter, which is the magnetic-field induced quantum phase transition 
to the magnon BEC phase, has been reported 
in $S = 1/2$ antiferromagnets with a spin gap, 
such as KCuCl$_{3}$, TlCuCl$_{3}$,\cite{Oosawa1999,PhysRevLett.84.5868,Tanaka2001,Yamada2008} 
BaCuSi$_{2}$O$_{6}$,\cite{Jaime2004} 
and Pb$_{2}$V$_{3}$O$_{9}$.\cite{Pb2V3O9}
The change of the magnon dispersion relation in TlCuCl$_{3}$ 
through the magnon BEC phase transition at $H_{\rm c} \sim 6$ T
has been observed by inelastic neutron scattering \cite{Ruegg2003}
and has been explained using the bond-operator representation.\cite{Matsumoto2004}
One of the characteristic features of the magnon BEC phase is 
the formation of massless excitation, i.e., the Goldstone mode 
at the magnetic $\Gamma$ point, 
indicating the spontaneous breaking of the continuous symmetry.
However the details of the magnon excitations, 
especially their symmetries, have not yet been established.
Raman scattering is a powerful tool 
to study phase transitions.
Because the magnon Raman process is sensitive 
to the symmetries of the ground and excited states,\cite{Fleury1968} 
Raman-scattering measurement above $H_{\rm c}$ presents great potential 
to study the change of the ground and excited states 
through the magnon BEC phase transition.

This paper reports the observation 
of one-magnon Raman scattering originating from 
changes of the ground and excited states 
through the magnon BEC phase transition.
This study focused on TlCuCl$_{3}$ 
where the magnon excitations and magnetic parameters below and above $H_{\rm c}$ 
have been studied in detail.\cite{Ruegg2003,Matsumoto2004} 
First, we show our experimental results above $H_{\rm c}$.
We then construct the microscopic theory 
of one-magnon Raman scattering 
in the exchange magnon Raman process
using the bond-operator representation, 
which can explain the experimental results clearly. 
Based on our results, the Raman selection rule will be clarified. 

\section{Experiments}

Single crystals of TlCuCl$_{3}$ were 
prepared by the vertical Bridgman method.\cite{Oosawa1999}
The 5145-\AA \  line of Ar$^+$-ion laser polarized along 
the (201) axis was incident on the (010) cleavage surface.
We set the samples in the cryostat under the dried N$_{2}$ or He gas atmosphere, 
because the sample was easily damaged by moisture in air.
We placed the microscope in the vacuum chamber of 
superconducting magnet in order to collect 
the scattered light effectively.
This enabled us to select good surface positions of 
crystals and the effects of the direct scattering 
in the low-energy region were avoided.
Magnetic fields of up to 10 T were applied 
nearly parallel to the $(010)$ axis.
The effect of a weak component of magnetic field along the (201) axis 
due to the experimental setting is negligible 
because the effect of the anisotropic $g$ tensor 
along these directions is small.\cite{Glazkov2004}

\section{Results}

\begin{figure}[t]
\includegraphics[width=0.45\textwidth]{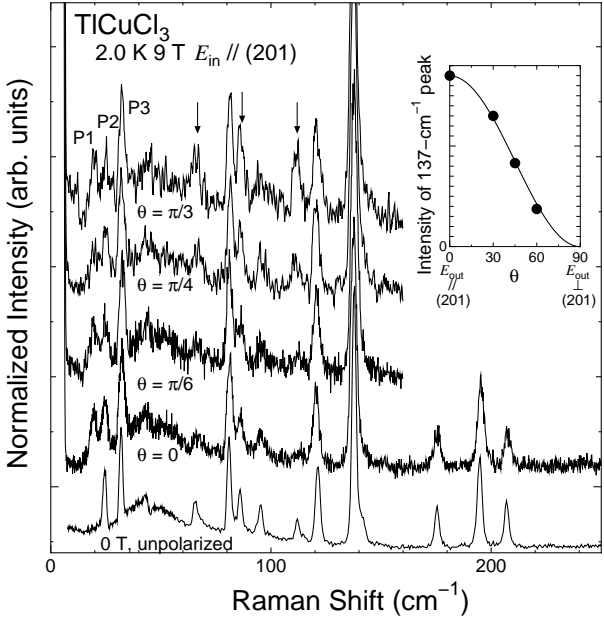}
\caption{
Polarization characteristics of Raman spectra at 9 T.
The inset shows the integrated Raman intensity of 
the phonon peak at 137 cm$^{-1}$.
The arrows indicates the Raman peaks 
coming from the $B_{\rm g}$ phonons.
The unpolarized Raman spectrum at 0 T is also shown.
}
\label{Polarization}
\end{figure}

Figure \ref{Polarization} compares 
the low-temperature Raman spectrum 
at 9 T (above $H_{\rm c}$ of TlCuCl$_{3}$) to that at 0 T.
At 0 T, we observed several sharp phonon peaks 
superimposed on the two-magnon Raman band 
extending from 11 cm$^{-1}$, 
i.e., twice the energy of the magnetic gap,\cite{Ruegg2003} 
 to about 150 cm$^{-1}$.
This spectrum is consistent with the results of refs. 
\onlinecite{Kusakabe2007} and \onlinecite{Choi2003}.
At 9 T, we observed the new Raman peak
with a Lorentzian lineshape, called P1, at 20 cm$^{-1}$.
No other significant change was observed.
P1 excitation has the $A_{\rm g}$ symmetry, 
which is obtained with the following procedures.
For the incident laser polarized along the (201) axis, 
${\bm E}_{\rm in} // (201)$, 
we measured the scattered light with polarization ${\bm E}_{\rm sc}$ 
which is rotated from ${\bm E}_{\rm in}$ with an angle $\theta$. 
The Raman intensities are normalized 
so that the 137-cm$^{-1}$ $A_{\rm g}$-symmetric phonon peaks in each spectrum, 
for which the $\theta$-dependence is shown in the inset of Fig. \ref{Polarization}, 
have the same intensity.
The Raman intensity 
from the quasiparticles with $A_{\rm g}$ symmetry, 
including P1, 
is $\theta$-independent in this plot 
while those with $B_{\rm g}$ symmetry indicated by arrows 
increased with increasing $\theta$, 
as shown in Fig. \ref{Polarization}.

\begin{figure}[t]
\includegraphics[width=0.45\textwidth]{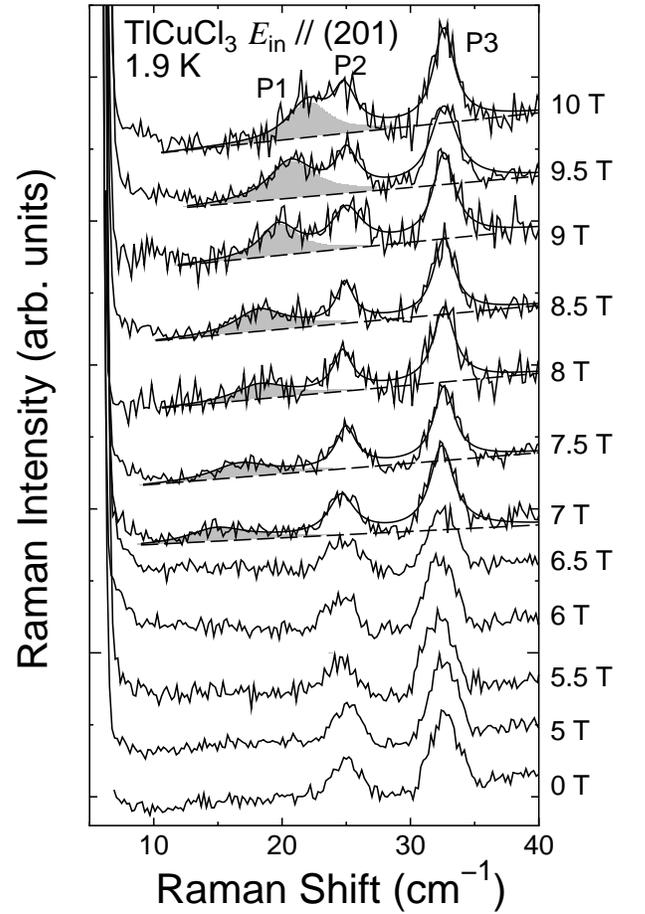}
\caption{
Magnetic-field dependence of Raman spectra 
in TlCuCl$_{3}$ at 1.9 K.
The fitting curves
are superimposed on the experimental data above 7 T.
The details of the fitting curve (solid curves) 
together with the background 
generated by the two-magnon Raman band (dashed lines)
are given in the text.
The hatched areas show the component of 
the Raman intensity generated by P1.
}
\label{RamanSpectra}
\end{figure}

Figure \ref{RamanSpectra} shows 
the detailed magnetic-field dependence of Raman spectra at 1.9 K.
At 0 T, the 25- and 32-cm$^{-1}$ phonon Raman peaks, 
called P2 and P3, respectively, 
are superimposed on the two-magnon Raman band starting at 11 cm$^{-1}$. 
P2 and P3 have the $A_{\rm g}$ symmetry as well as P1, 
as shown in Fig. \ref{Polarization}.
The Raman spectra below 5 T are magnetic-field independent.
Above 7 T, we clearly observed that 
the frequency and intensity of P1 strongly depended 
on the applied magnetic field, 
as denoted by the hatched areas in Fig. \ref{RamanSpectra}, 
of which the details will be explained later.
Around 6 T, 
the increase of the Rayleigh scattering around 0 cm$^{-1}$ 
suggests the quasielastic (or critical) light scattering 
reflecting the large magnetic specific heat around $H_{\rm c}$ 
as observed in several antiferromagnets 
or spin-Peierls system.\cite{FePS3,Kuroe1997,Halley1978}
To discuss the quasielastic light scattering quantitatively, 
Raman-scattering measurements in the anti-Stokes region are necessary.

The lineshapes of P1, P2, and P3 
are well described by three Lorentzian curves 
superimposed on the background: 
\begin{equation}
\displaystyle
I(\omega) = 
\sum_{i = 1}^{3}
\frac{(n + 1)k_{i}^{2} \omega \Gamma_{i}}{( \omega^{2} - \omega_{i}^{2} )^{2} 
+ (\omega \Gamma_{i})^{2}}
+ {\rm background}
\ , 
\label{Lorentzian}
\end{equation}
where $k_{i}$, $\hbar\omega_{i}$, and $\Gamma_{i}$ indicate 
the Raman coupling coefficient, the energy, 
and the halfwidth of P$i$, respectively.
Here, the Bose factor $(n + 1)$ can be treated as unity 
because the temperature is much lower than the energies of quasiparticles.
The background generated by the two-magnon Raman band 
peaking around 50 cm$^{-1}$
was assumed to be a linear function, 
as shown 
by the dashed lines in Fig. \ref{RamanSpectra}.
The calculated curves reproduced the observed data well.
We show the Raman intensity 
generated by P1 (the term related to the subscript $i$ = 1 in eq. (\ref{Lorentzian})) 
on the linear background 
as the hatched area in Fig. \ref{RamanSpectra}.
Around $H_{\rm c}$, 
we could not distinguish P1 from 
the two-magnon Raman band
because of its weak intensity.

\begin{figure}[t]
\includegraphics[width=0.45\textwidth]{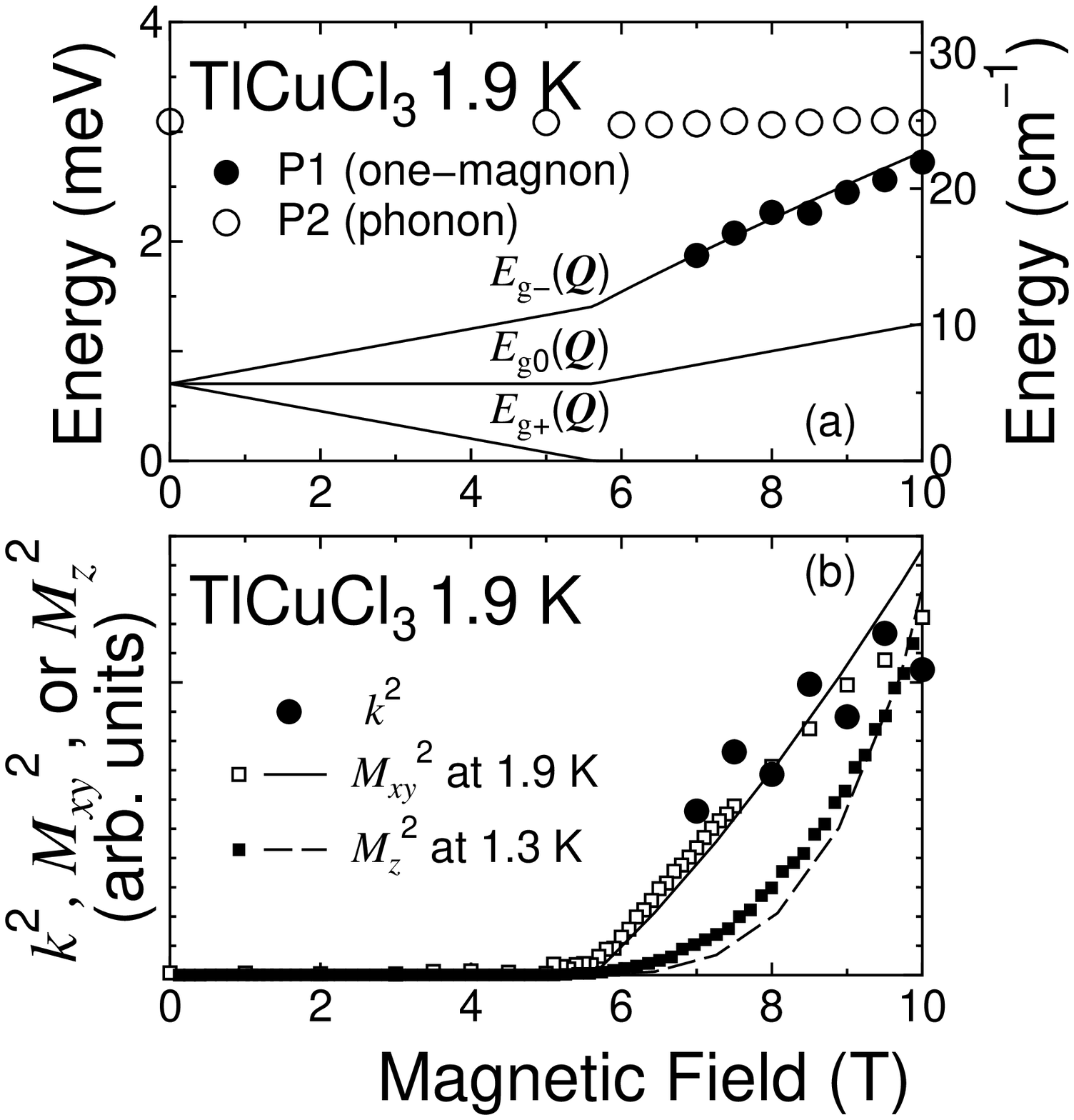}
\caption{
(a) Magnetic-field dependence of one-magnon (filled circles) 
and phonon (empty circles) energies 
together with that of one-magnon energies
calculated in refs. 
\onlinecite{Matsumoto2004} and \onlinecite{PhysRevLett.89.077203}.
(b) Magnetic-field dependence of 
squared Raman coupling coefficient (filled circles).
Squared transverse magnetization $M_{xy}^{2}$ (empty squares, ref. \onlinecite{Tanaka2001}) 
and squared longitudinal magnetization $M_{z}^{2}$ 
(filled squares, ref. \onlinecite{Takatsu1997}) 
are also plotted with the calculated values for them 
(a solid and a dashed curves 
for $M_{xy}^{2}$ and $M_{z}^{2}$, respectively).\cite{Matsumoto2004}
}
\label{RamanParms}
\end{figure}

Figure \ref{RamanParms}(a) shows 
the peak energies $\hbar\omega_{1}$ and $\hbar\omega_{2}$ 
as functions of magnetic field 
together with the calculated one-magnon energy $E_{g\alpha}({\bm Q})$ 
($\alpha = -,0,+$) with the wavevector $\bm{Q} = (0,0,2\pi)$,
\cite{PhysRevLett.89.077203,Matsumoto2004} 
where the magnetic gap is closed.\cite{Tanaka2001,Oosawa2002} 
One can see that $\hbar \omega_{1}$ below 10 T agrees with $E_{g-}({\bm Q})$ 
within experimental accuracy. 
Figure \ref{RamanParms}(b) shows the squared Raman coupling coefficient 
$k^{2} = k_{1}^{2}/k_{3}^{2}$, 
which is proportional to the integrated Raman intensity of P1.
Here, we normalized $k_{1}^{2}$, 
which is proportional to the area hatched in Fig. \ref{RamanSpectra}, 
by $k_{3}^{2}$ 
to correct errors due to the small deviations of optic alignment.
The errors of $k^{2}$ are similar to the symbol size in Fig. \ref{RamanParms}(b).
For comparison, we show the magnetic-field dependences of 
squared transverse magnetization $M_{xy}^{2}$ (ref. \onlinecite{Tanaka2001})
and squared longitudinal magnetization  $M_{z}^{2}$ (ref. \onlinecite{Takatsu1997})
together with their calculated values.\cite{Matsumoto2004}
One can see that the magnetic-field dependence 
of Raman intensity is well scaled to the former.

\section{\label{Discussion}Discussion}

First, we discuss the origin of P1.
Judging from the polarization characteristics, 
the scattering process of P1 comes from 
the exchange magnon Raman scattering, as proposed by Fleury and Loudon,\cite{Fleury1968} 
which usually creates the broad two-magnon Raman band
with $A_{\rm g}$ symmetry reported at zero magnetic field.\cite{Choi2003,Kusakabe2007} 
Because the two-magnon Raman band at 9 T 
is almost the same as that at 0 T, as seen in Fig. \ref{Polarization},
we do not need to consider 
the increase of the magnon-magnon interaction 
which may cause the two-magnon Raman band 
with a nearly Lorentzian lineshape \cite{Weber1989,PhysRevB.42.4842} 
as well as the formation of the two-magnon bound state.\cite{Sekine1998}
P1 does not originate from a three-magnon process, 
where the thermally excited triplets play 
an essential role.\cite{PhysRevLett.79.5138,Choi2005PRB}
The three-magnon Raman intensity, 
which is proportional to $n$ at low temperatures, 
should be negligibly small at 1.9 K.
Because P1 has a Lorentzian lineshape, 
the one-magnon Raman scattering can be considered 
as the origin of P1.
Hereafter, 
we consider the detail of the magnon Raman process 
using the bond-operator representation 
and clarify that the one-magnon Raman scattering 
from the exchange magnon Raman process becomes possible 
in the magnon BEC phase.

The effective Raman operator ${\cal R}$ 
in the above-mentioned exchange magnon Raman process 
has a form given by the isotropic 
Heisenberg-type exchange interaction 
between spins $\bm{S}_{i}$ and $\bm{S}_{j}$:
\begin{equation}
{\cal R} = 
\sum_{i,j} {\cal R}_{i,j} = 
\sum_{i,j} F_{i,j} 
(\hat{\bm E}_{\rm in} \cdot \hat{\bm r}_{ij})
(\hat{\bm E}_{\rm sc} \cdot \hat{\bm r}_{ij}) 
\bm{S}_{i} \cdot \bm{S}_{j} \ , 
\label{FleuryLoudon}
\end{equation}
where $\bm{r}_{ij}$ indicates the position vector 
between $\bm{S}_{i}$ and $\bm{S}_{j}$ 
and the sum runs over all the interacting spin pairs.
Here, 
$\hat{\bm r} = {\bm r}/|{\bm r}|$, 
$\hat{\bm E} = {\bm E}/|{\bm E}|$, 
and the coefficient $F_{i,j}$ depends on the pathway of interaction 
between $\bm{S}_{i}$ and $\bm{S}_{j}$.
${\cal R}$ depends on the experimental setting
through the 
$(\hat{\bm E}_{\rm in} \!\! \cdot \! \hat{\bm r}_{ij})
 (\hat{\bm E}_{\rm sc} \!\! \cdot \! \hat{\bm r}_{ij})$ term. 
The matrix element of ${\cal R}$ between the initial state $| {\rm i} \rangle$
and the final one $| {\rm f} \rangle$
is called Raman tensor.
The magnon Raman intensity is given as 
\begin{equation}
{\cal I}^{\rm (in,sc)} (\omega) \propto
\displaystyle
|\bm{E}_{\rm in}|^{2}
\sum_{| {\rm f} \rangle} 
\left|
\langle {\rm i} | {\cal R} | {\rm f} \rangle
\right|^{2}
\delta(\omega - \omega_{\rm if})
\ , 
\end{equation}
where $\hbar \omega_{\rm if}$ is the 
excitation energy between the states $|{\rm i}\rangle$ and $|{\rm f}\rangle$.

We note here that 
${\cal R}$ is {\it always} written 
using the pure singlet operator $s_{\bm{k}}$ 
and the triplet operators $t_{\bm{k}\alpha}$ $(\alpha = -,0,+)$, 
which annihilate 
the triplets with $S^{z} = \alpha$.
We need to rewrite ${\cal R}$  
using the creation and annihilation operators 
of eigenstates in the magnon BEC phase.
As discussed by Matsumoto {\it et al.},
\cite{PhysRevLett.89.077203,Matsumoto2004} 
the following transformed operators 
based on the bond-operator representation 
characterize the magnon excitations in the magnon BEC phase:
\begin{equation}
\begin{array}{l}
\displaystyle
s_{\bm{k}} \! = \! u a_{\bm{k}} \! - \! v b_{\bm{k}+\bm{Q}+}, 
\\
\displaystyle
t_{\bm{k}+}  \! = \!  vf a_{\bm{k}-\bm{Q}} \! + \! uf b_{\bm{k}+} \! - \! g b_{\bm{k}-} , 
\\ 
\displaystyle
t_{\bm{k}0}  \! = \!   b_{\bm{k}0}, 
\\
\displaystyle
t_{\bm{k}-} \! = \!  vg a_{\bm{k}-\bm{Q}} \! + \! ug b_{\bm{k}+} \! + \! f b_{\bm{k}-} 
, 
\label{mixedrepresentation}
\end{array}
\end{equation}
where the momentum-independent real-number parameters
$u$, $v$, $f$, and  $g$ satisfy $f^{2}+g^{2} = u^{2}+v^{2} = 1$.
Below $H_{\rm c}$, $v = 0$ 
and one can obtain the simple relations
$a_{\bm{k}} = s_{\bm{k}}$ and $b_{\bm{k}\alpha} = t_{\bm{k}\alpha}$.
Above $H_{\rm c}$, 
$v \neq 0$ 
and the operators $a_{\bm{k}-\bm{Q}}$ and $b_{\bm{k}\pm}$ are linearly combined.
The mixing of $b_{\bm{k}\pm}$ and $b^{\dagger}_{-\bm{k}\pm}$ 
is treated using the Bogoliubov transformation, 
as will be shown in detail later.
It should be noted that 
the ground state does not include 
the $b_{\bm{k}0}$ state, 
indicating that the $E_{g0}({\bm Q})$ mode is not Raman-active.

We consider the Raman operator ${\cal R}_{\bm d}$ 
associated with the intradimer interaction, 
which  can be written in the reciprocal lattice space as 
\begin{equation}
\begin{array}{l}
{\cal R}_{\bm d} 
 = 
\displaystyle
F_{\bm d} 
(\hat{\bm{E}_{\rm in}} \!\! \cdot \! \hat{\bm{d}})
(\hat{\bm{E}_{\rm sc}} \!\! \cdot \! \hat{\bm{d}})
\!\!
\sum_{\bm{k}} 
  \!\!
  \left( \!\!
    - \frac{3}{4} 
  s_{\bm{k}}^{\dagger} s_{\bm{k}} 
+ \sum_{\alpha} \!
    \frac{1}{4}
    t_{\bm{k}\alpha}^{\dagger}t_{\bm{k}\alpha}
\!\!
\right)
\\
=
\displaystyle
F_{\bm d} 
(\hat{\bm{E}_{\rm in}} \!\! \cdot \! \hat{\bm{d}})
(\hat{\bm{E}_{\rm sc}} \!\! \cdot \! \hat{\bm{d}})
\! \left[ \!
\left( \! \frac{1}{4}-u^{2} \! \right) \! \bar{a}^{2}
\! + \! uv\bar{a}(b_{\bm{Q}+}+b_{\bm{Q}+}^{\dagger})
\right.
\\
\displaystyle
+
\left.
\sum_{{\bm k}}
\left\{
\!\! \left( \! \frac{1}{4} - v^{2} \! \right) \!\! b_{\bm{k}+}^{\dagger}b_{\bm{k}+}
+ \frac{1}{4} b_{\bm{k}-}^{\dagger}b_{\bm{k}-}
+ \frac{1}{4} b_{\bm{k}0}^{\dagger}b_{\bm{k}0}
\right\}
\! \right]
\ ,
\end{array}
\label{Rintra}
\end{equation}
where ${\bm d}$ indicates 
the position vector 
between two spins forming a dimer, 
which is almost parallel to the $(201)$ direction, 
and $F_{\bm d}$ originates from $F_{i,j}$.
Here, we used the facts that 
the operator $a_{\bm{k}}$ can be treated as 
a uniformly condensed mean-field parameter $\bar{a}\delta_{\bm{k},0}$ 
(ref. \onlinecite{PhysRevLett.89.077203}) and 
$b_{\bm{-Q}+}^{\dagger} = b_{\bm{Q}+}^{\dagger}$ at the magnetic $\Gamma$ point.
The term $uv\bar{a}(b_{\bm{Q}+} + b_{\bm{Q}+}^{\dagger})$ in eq. (\ref{Rintra}) 
gives the momentum selection rule of one-magnon Raman scattering.
The parameter $uv$ indicates 
the appearance of one-magnon Raman scattering only above $H_{\rm c}$ and 
the Raman intensity 
proportional to $M_{xy}^{2}$ (see eq. (5) of ref. \onlinecite{PhysRevLett.89.077203}), 
which is consistent with the observation, 
as shown in Fig. \ref{RamanParms}(b).
The terms $\bar{a}^{2}$ and $b_{{\bm k}\alpha}^{\dagger}b_{{\bm k}\alpha}$ 
in eq. (\ref{Rintra})
do not give one-magnon Raman scattering.

The quadratic terms of the magnetic Hamiltonian ${\cal H}_{\pm}$ in ref. \onlinecite{Matsumoto2004} 
can be diagonalized using the $\alpha_{\bm k}^{\pm}$ bosonic operators 
which annihilate the $E_{g\pm}({\bm k})$ modes, respectively.
The ground state in the magnon BEC phase is 
the vacuum state for the $\alpha_{\bm k}^{\pm}$ operators.
Using the bosonic commutation relations of $\alpha_{\bm k}^{\pm}$,
we obtain the inverse Bogoliubov transformation as 
\begin{equation}
\left( \!\!
\begin{array}{c}
b_{\bm k-} \\
b_{\bm k+} \\
b_{-{\bm k}-}^{\dagger} \\
b_{-{\bm k}+}^{\dagger} 
\end{array}
\!\! \right)
=
\left(
\!\!
\begin{array}{cccc}
\! u_{-{\bm k}-}^{-*} & 
\!\! u_{-{\bm k}-}^{+*} &
\!\!\!-v_{{\bm k}-}^{-} &
\!\!-v_{{\bm k}-}^{+} \\
\! u_{-{\bm k}+}^{-*} & 
\!\! u_{-{\bm k}+}^{+*} &
\!\!\!-v_{{\bm k}+}^{-} &
\!\!-v_{{\bm k}+}^{+} \\
\!-v_{-{\bm k}-}^{-*} &
\!\!-v_{-{\bm k}-}^{+*} &
\!\!\! u_{{\bm k}-}^{-} & 
\!\! u_{{\bm k}-}^{+} \\
\!-v_{-{\bm k}+}^{-*} &
\!\!-v_{-{\bm k}+}^{+*} &
\!\!\! u_{{\bm k}+}^{-} & 
\!\! u_{{\bm k}+}^{+} \\
\end{array}
\!\!
\right)
\!\!\!
\left(
\!\!
\begin{array}{c}
\alpha_{\bm k}^{-}\\
\alpha_{\bm k}^{+}\\
\alpha_{-{\bm k}}^{-\dagger}\\
\alpha_{-{\bm k}}^{+\dagger}
\end{array}
\!\!
\right)
\ .
\label{InvTransformation}
\end{equation}
\begin{figure}[t]
\includegraphics[width=0.45\textwidth]{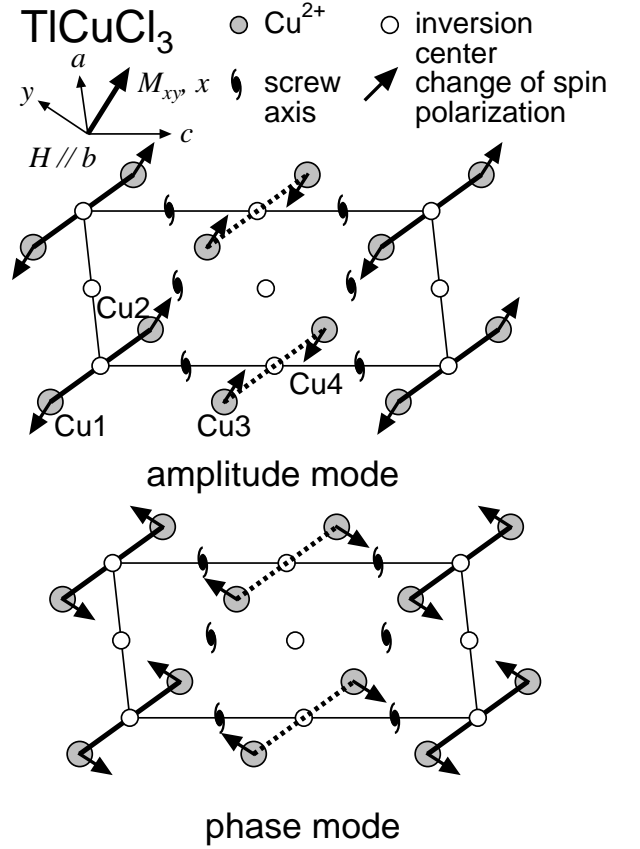}
\caption{Changes of spin polarizations for 
the amplitude and phase modes, 
corresponding to the ionic-vibration patterns in phonons.
The solid and dashed lines indicate the dimers 
at the corner and center of the chemical unit cell 
denoted by parallelograms, respectively.
Only the Cu$^{2+}$ sites are shown 
with the symbols of inversion centers and screw axes.
The direction of the transverse magnetization $x$ 
in the magnon BEC phase is shown by 
a bold arrow with $M_{xy}$.\cite{Tanaka2001}
The direction $y$ is perpendicular to $x$ and $b$.}
\label{Struct}
\end{figure}
The one-magnon term in eq. (\ref{Rintra}) can be rewritten as 
\begin{equation}
b_{\bm Q+} + b_{{\bm Q}+}^{\dagger} 
\!=\! 
(u_{Q+}^{-}-v_{Q+}^{-})^{\!*}\!\alpha_{\bm Q}^{-}
+ (u_{Q+}^{+}-v_{Q+}^{+})^{\!*}\!\alpha_{\bm Q}^{+}
+ {\rm H. c.} \ ,
\end{equation}
indicating that both the $E_{g+}({\bm Q})$ and $E_{g-}({\bm Q})$ modes 
are symmetry-allowed.
These modes can be described as 
the mixed amplitude and phase modes 
which are related to 
the spin correlation functions 
along the $x$ and $y$ directions in Fig. \ref{Struct}, 
respectively.
The amplitude mode changes the amplitudes of $M_{xy}$ 
without changing their directions 
whereas the phase mode is described as uniform rotations of $M_{xy}$.
Both of these modes have $A_{\rm g}$ symmetry, 
i.e., these are intrinsically Raman-active 
because the continuous {\it rotational} symmetry is broken 
above $H_{\rm c}$.
This is one of the most distinguishing characteristics 
of the magnetic excitations in the magnon BEC phase.
In systems of density waves,\cite{Gruner1994} 
the Goldstone mode (phase mode), 
which corresponds to the continuous {\it translational} operation, 
is IR-active.

Let us show that 
the phase mode does not give the finite Raman intensity.
When the phase mode is the Goldstone mode, i.e., for $E_{g+}({\bm Q}) = 0$,  
${\cal H}_{\pm}$ can be diagonalized by 
\begin{equation}
\left(
\!\!\!
\begin{array}{cccc}
\epsilon_{{\bm Q}+}      &-\Delta_{{\bm Q}+}         & \epsilon_{{\bm Q}\pm}    &-\Delta_{{\bm Q}\pm}     \\
\Delta_{{\bm Q}+}        &-\epsilon_{{\bm Q}+}       & \Delta_{{\bm Q}\pm}      &-\epsilon_{{\bm Q}\pm}   \\
\epsilon_{{\bm Q}\pm}    &-\Delta_{{\bm Q}\pm}       & \epsilon_{{\bm Q}-}      &-\Delta_{{\bm Q}-}       \\
\Delta_{{\bm Q}\pm}      &-\epsilon_{{\bm Q}\pm}     & \Delta_{{\bm Q}-}        &-\epsilon_{{\bm Q}-}   
\end{array}
\!\!\!
\right)
\!\!\!
\left( \!\!
\begin{array}{c}
u_{{\bm Q}+}^{+}\\
v_{{\bm Q}+}^{+}\\
u_{{\bm Q}-}^{+}\\
v_{{\bm Q}-}^{+}\\
\end{array}
\!\!\! \right) = 0, 
\label{HamiltonianMatrix}
\end{equation}
where the definitions of $\epsilon_{{\bm Q}\beta}$ and $\Delta_{{\bm Q}\beta}$
($\beta = +, -, \pm$) are given in ref. \onlinecite{Matsumoto2004}.
We can reduce eq. (\ref{HamiltonianMatrix}) to the form of 
\begin{equation}
\left(
\!\!\!
\begin{array}{cc}
\epsilon_{{\bm Q}+}   + \Delta_{{\bm Q}+}     & \epsilon_{{\bm Q}\pm}    + \Delta_{{\bm Q}\pm}     \\
\epsilon_{{\bm Q}\pm} + \Delta_{{\bm Q}\pm}   & \epsilon_{{\bm Q}-}      + \Delta_{{\bm Q}-}       
\end{array}
\!\!\!
\right)
\!\!\!
\left( \!\!
\begin{array}{c}
u_{{\bm Q}+}^{+} - 
v_{{\bm Q}+}^{+}\\
u_{{\bm Q}-}^{+} -
v_{{\bm Q}-}^{+} 
\end{array}
\!\!\! \right) = 0.
\label{HamiltonianMatrixSmall}
\end{equation}
Because the matrix in eq. (\ref{HamiltonianMatrixSmall}) is invertible, 
we find that $u_{{\bm Q}+}^{+} - v_{{\bm Q}+}^{+} = 0$, 
i.e., the Raman intensity for the phase mode is zero 
although this mode is symmetry-allowed.
This result indicates that 
only the spin correlation function along $x$ is detectable 
with a factor $M_{xy}^{2}$ by the first-order Raman scattering.

In case of TlCuCl$_{3}$, 
the anisotropic exchange interaction gives a small magnetic gap 
$E_{g+}({\bm Q}) \approx 1.7$ cm$^{-1}$.\cite{Glazkov2004} 
In this case, the $E_{g+}({\bm Q})$ mode 
is the mixed amplitude and phase modes and 
it gives a finite one-magnon Raman intensity.
In our measurements, however, 
we could not detect it 
because of the strong direct scattering at 0 cm$^{-1}$.
It is worthwhile to consider 
the Raman scattering from the $E_{g+}({\bm Q})$ 
because this mode is thermally populated at 1.9 K.
The transition to the ground state 
is the one-magnon anti-Stokes Raman scattering, 
which cannot be detected in our measurements, as stated above.
The transition to the $E_{g-}({\bm Q})$ mode is obtained 
from the terms $\alpha_{\bm k}^{-\dagger}\alpha_{\bm k}^{+}$ in Raman tensor.
Substituting $\alpha_{\bm k}^{\pm}$ and $\alpha_{\bm k}^{\pm\dagger}$ 
in eq. (\ref{InvTransformation}) 
for $b_{\bm k\pm}$ and $b_{\bm k\pm}^{\dagger}$ in eq. (\ref{Rintra}), 
${\cal R}_{\bm d}$ contains the terms 
\begin{equation}
\begin{array}{l}
\displaystyle
\sum_{\bm k}
\left\{
\!\frac{1}{4}\!\left(
u_{-{\bm k}+}^{+*}\!u_{-{\bm k}+}^{-}\!\!+\!
v_{-{\bm k}+}^{+*}\!v_{-{\bm k}+}^{-}\!\!+\!
u_{-{\bm k}-}^{+*}\!u_{-{\bm k}-}^{-}\!\!+\!
v_{-{\bm k}-}^{+*}\!v_{-{\bm k}-}^{-}
\right)
\right.
\\ 
\displaystyle
\ \ \ \ \ \ \ 
\biggl.
\!-v^{2}\!\left(
u_{-{\bm k}+}^{+*}\!u_{-{\bm k}+}^{-}\!\!+\!
v_{-{\bm k}+}^{+*}\!v_{-{\bm k}+}^{-}
\right)
\!\biggr\}
\alpha_{\bm k}^{-\dagger}\alpha_{\bm k}^{+} 
\ .
\end{array}
\end{equation}
This result indicates that 
the transition from the $E_{g+}({\bm Q})$ state 
to the $E_{g-}({\bm Q})$ one may be detected 
as a part of the two-magnon Raman band 
and its intensity is expected to have no drastic change, 
at least below 10 T, because $v^{2}$ below 10 T is 
very small.\cite{Matsumoto2004,PhysRevLett.89.077203}
Actually, the profile and intensity of the two-magnon Raman band 
at 9 T are almost similar to those at 0 T, as shown in Fig. \ref{Polarization}.

When we consider one-magnon Raman scattering 
caused by the interdimer interaction, 
we can substitute the expectation value 
for one of the spin operators in eq. (\ref{FleuryLoudon}):
\begin{equation}
{\cal R} = 
\sum_{i,j} F_{i,j} 
(\hat{\bm{E}_{\rm in}} \cdot \hat{\bm{r}_{ij}})
(\hat{\bm{E}_{\rm sc}} \cdot \hat{\bm{r}_{ij}}) 
\bm{S}_{i} \cdot \langle \bm{S}_{j} \rangle \ .
\label{FleuryLoudonMF}
\end{equation}
The Raman tensor from the interdimer interaction 
also contains the terms 
$(b_{\bm{Q}+} + b_{\bm{Q}+}^{\dagger})$, suggesting that 
the one-magnon Raman scattering from the $E_{g-}(\bm{Q})$ mode 
can be detected above $H_{\rm c}$.
Because $\langle S \rangle = M_{xy}$, 
one can expect that 
the Raman intensity is also proportional to $M_{xy}^{2}$.
The precise analytic form of the Raman tensor
coming from the interdimer interaction has been established
in our recent letter in case of the pressure-induced
magnon BEC phase at zero magnetic field. \cite{Matsumoto2008}
The magnetic field-induced magnon BEC case
will be published elsewhere. \cite{MatsumotoFuture}

We point out that our theory 
for the appearance of the one-magnon Raman scattering is applicable to 
the pressure-induced magnon BEC phase transition in TlCuCl$_{3}$.
\cite{Matsumoto2004,Goto2004,Oosawa2003}
In case of pressure-induced magnon BEC phase transition, 
the pure amplitude mode, 
which is the longitudinal spin-wave mode 
coupled only with the spin correlation function 
along the $x$ direction in Fig. \ref{Struct},
is expected to be observed.\cite{Matsumoto2007,Matsumoto2008}

The two-magnon Raman band is also interesting 
as well as the appearance of the one-magnon Raman peak
which is the main purpose of this paper.
In case of two-magnon Raman scattering, 
both ${\cal R}_{\bm d}$ and 
the Raman tensors generated from 
the interdimer interactions, 
which create the magnon pair with the zero total momentum, 
play an essential role.
At present, it was difficult to calculate 
the lineshape of the two-magnon Raman band 
because the values of $F_{i,j}$, 
which the two-magnon Raman spectrum is sensitive to, 
cannot be obtained directly.

\section{Conclusion}

We have assigned 
the origin of the Raman peak appearing above $H_{\rm c}$ in TlCuCl$_{3}$
to one-magnon Raman scattering, 
which comes from the exchange magnon Raman process. 
This is based on 
(1) the Lorentzian lineshape of the peak, 
(2) its Raman shift tracing $E_{g-}({\bm Q})$, 
(3) its polarization characteristics, 
i.e. this one-magnon Raman scattering is $A_{\rm g}$-symmetric  
as well as the second-order magnetic Raman scattering, 
and (4) the observation that the peak's Raman intensity is proportional to $M_{xy}^{2}$.
Using the bond-operator representation, 
we calculated the Raman intensity 
to clarify the Raman selection rule 
of one-magnon Raman scattering in the exchange magnon Raman process. 
The intensity of the one-magnon Raman scattering is 
related to the spin correlation function along the direction of $M_{xy}$, 
i.e., the $x$ direction in Fig. 4.
And therefore the $E_{g\pm}({\bm Q})$ modes with the finite excitation energies 
are $A_{\rm g}$-symmetric and Raman-active.
In the isotropic limit, 
the Goldstone mode for $E_{g+}({\bm Q}) = 0$, 
which is related to the spin correlation function along the $y$ direction, 
is $A_{\rm g}$-symmetric but has no Raman intensity.
The $E_{g0}({\bm Q})$ mode and 
the magnetic excitation at the chemical $\Gamma$ point are Raman-inactive. 
The change of the ground and excited states 
through the magnon BEC phase transition can be detected 
via the appearance of a new one-magnon Raman peak 
from the $E_{g-}({\bm Q})$ mode in the magnon BEC phase.


\begin{thebibliography}{29}
\expandafter\ifx\csname natexlab\endcsname\relax\def\natexlab#1{#1}\fi
\expandafter\ifx\csname bibnamefont\endcsname\relax
  \def\bibnamefont#1{#1}\fi
\expandafter\ifx\csname bibfnamefont\endcsname\relax
  \def\bibfnamefont#1{#1}\fi
\expandafter\ifx\csname citenamefont\endcsname\relax
  \def\citenamefont#1{#1}\fi
\expandafter\ifx\csname url\endcsname\relax
  \def\url#1{\texttt{#1}}\fi
\expandafter\ifx\csname urlprefix\endcsname\relax\def\urlprefix{URL }\fi
\providecommand{\bibinfo}[2]{#2}
\providecommand{\eprint}[2][]{\url{#2}}

\bibitem[{\citenamefont{Oosawa et~al.}(1999)\citenamefont{Oosawa, Ishii, and
  Tanaka}}]{Oosawa1999}
\bibinfo{author}{\bibfnamefont{A.}~\bibnamefont{Oosawa}},
  \bibinfo{author}{\bibfnamefont{M.}~\bibnamefont{Ishii}}, \bibnamefont{and}
  \bibinfo{author}{\bibfnamefont{H.}~\bibnamefont{Tanaka}},
  \bibinfo{journal}{J. Phys. C: Condens. Matter} \textbf{\bibinfo{volume}{11}},
  \bibinfo{pages}{265} (\bibinfo{year}{1999}).

\bibitem[{\citenamefont{Nikuni et~al.}(2000)\citenamefont{Nikuni, Oshikawa,
  Oosawa, and Tanaka}}]{PhysRevLett.84.5868}
\bibinfo{author}{\bibfnamefont{T.}~\bibnamefont{Nikuni}},
  \bibinfo{author}{\bibfnamefont{M.}~\bibnamefont{Oshikawa}},
  \bibinfo{author}{\bibfnamefont{A.}~\bibnamefont{Oosawa}}, \bibnamefont{and}
  \bibinfo{author}{\bibfnamefont{H.}~\bibnamefont{Tanaka}},
  \bibinfo{journal}{Phys. Rev. Lett.} \textbf{\bibinfo{volume}{84}},
  \bibinfo{pages}{5868} (\bibinfo{year}{2000}).

\bibitem[{\citenamefont{Tanaka et~al.}(2001)\citenamefont{Tanaka, Oosawa, Kato,
  Uekusa, Ohashi, Kakurai, and Hoser}}]{Tanaka2001}
\bibinfo{author}{\bibfnamefont{H.}~\bibnamefont{Tanaka}},
  \bibinfo{author}{\bibfnamefont{A.}~\bibnamefont{Oosawa}},
  \bibinfo{author}{\bibfnamefont{T.}~\bibnamefont{Kato}},
  \bibinfo{author}{\bibfnamefont{H.}~\bibnamefont{Uekusa}},
  \bibinfo{author}{\bibfnamefont{Y.}~\bibnamefont{Ohashi}},
  \bibinfo{author}{\bibfnamefont{K.}~\bibnamefont{Kakurai}}, \bibnamefont{and}
  \bibinfo{author}{\bibfnamefont{A.}~\bibnamefont{Hoser}}, \bibinfo{journal}{J.
  Phys.\ Soc.\ Japan.} \textbf{\bibinfo{volume}{70}}, \bibinfo{pages}{939}
  (\bibinfo{year}{2001}).

\bibitem[{\citenamefont{Yamada et~al.}(2008)\citenamefont{Yamada, Ono, Tanaka,
  Misguich, Oshikawa, and Sakakibara}}]{Yamada2008}
\bibinfo{author}{\bibfnamefont{F.}~\bibnamefont{Yamada}},
  \bibinfo{author}{\bibfnamefont{T.}~\bibnamefont{Ono}},
  \bibinfo{author}{\bibfnamefont{H.}~\bibnamefont{Tanaka}},
  \bibinfo{author}{\bibfnamefont{G.}~\bibnamefont{Misguich}},
  \bibinfo{author}{\bibfnamefont{M.}~\bibnamefont{Oshikawa}}, \bibnamefont{and}
  \bibinfo{author}{\bibfnamefont{T.}~\bibnamefont{Sakakibara}},
  \bibinfo{journal}{J.\ Phys.\ Soc.\ Jpn.} \textbf{\bibinfo{volume}{77}},
  \bibinfo{pages}{013701} (\bibinfo{year}{2008}).

\bibitem[{\citenamefont{Jaime et~al.}(2004)\citenamefont{Jaime, Correa,
  Harrison, Batista, Kawashima, Kazuma, Jorge, Stein, Heinmaa, Zvyagin
  et~al.}}]{Jaime2004}
\bibinfo{author}{\bibfnamefont{M.}~\bibnamefont{Jaime}},
  \bibinfo{author}{\bibfnamefont{V.~F.} \bibnamefont{Correa}},
  \bibinfo{author}{\bibfnamefont{N.}~\bibnamefont{Harrison}},
  \bibinfo{author}{\bibfnamefont{C.~D.} \bibnamefont{Batista}},
  \bibinfo{author}{\bibfnamefont{N.}~\bibnamefont{Kawashima}},
  \bibinfo{author}{\bibfnamefont{Y.}~\bibnamefont{Kazuma}},
  \bibinfo{author}{\bibfnamefont{G.~A.} \bibnamefont{Jorge}},
  \bibinfo{author}{\bibfnamefont{R.}~\bibnamefont{Stein}},
  \bibinfo{author}{\bibfnamefont{I.}~\bibnamefont{Heinmaa}},
  \bibinfo{author}{\bibfnamefont{S.~A.} \bibnamefont{Zvyagin}},
  \bibinfo{author}{\bibfnamefont{Y.}~\bibnamefont{Sasago}}, \bibnamefont{and}
  \bibinfo{author}{\bibfnamefont{K.}~\bibnamefont{Uchinokura}},
  \bibinfo{journal}{Phys.\ Rev.\ Lett.} \textbf{\bibinfo{volume}{93}},
  \bibinfo{pages}{087203} (\bibinfo{year}{2004}).

\bibitem[{\citenamefont{Waki et~al.}(2004)\citenamefont{Waki, Morimoto,
  Michioka, Kato, Kageyama, Yoshimura, Nakatsuji, Sakai, Maeno, Mitamura
  et~al.}}]{Pb2V3O9}
\bibinfo{author}{\bibfnamefont{T.}~\bibnamefont{Waki}},
  \bibinfo{author}{\bibfnamefont{Y.}~\bibnamefont{Morimoto}},
  \bibinfo{author}{\bibfnamefont{C.}~\bibnamefont{Michioka}},
  \bibinfo{author}{\bibfnamefont{M.}~\bibnamefont{Kato}},
  \bibinfo{author}{\bibfnamefont{H.}~\bibnamefont{Kageyama}},
  \bibinfo{author}{\bibfnamefont{K.}~\bibnamefont{Yoshimura}},
  \bibinfo{author}{\bibfnamefont{S.}~\bibnamefont{Nakatsuji}},
  \bibinfo{author}{\bibfnamefont{O.}~\bibnamefont{Sakai}},
  \bibinfo{author}{\bibfnamefont{Y.}~\bibnamefont{Maeno}},
  \bibinfo{author}{\bibfnamefont{H.}~\bibnamefont{Mitamura}}, \bibnamefont{and}
  \bibinfo{author}{\bibfnamefont{T.}~\bibnamefont{Goto}}, \bibinfo{journal}{J.\
  Phys.\ Soc.\ Jpn.} \textbf{\bibinfo{volume}{73}}, \bibinfo{pages}{3435}
  (\bibinfo{year}{2004}).

\bibitem[{\citenamefont{R{\"u}egg et~al.}(2003)\citenamefont{R{\"u}egg,
  Cavadini, G{\"u}del, Kr{\"a}mer, Mutka, Wildes, Habicht, and
  Vorderwisch}}]{Ruegg2003}
\bibinfo{author}{\bibfnamefont{C.}~\bibnamefont{R{\"u}egg}},
  \bibinfo{author}{\bibfnamefont{N.}~\bibnamefont{Cavadini}},
  \bibinfo{author}{\bibfnamefont{H.-U.} \bibnamefont{G{\"u}del}},
  \bibinfo{author}{\bibfnamefont{K.}~\bibnamefont{Kr{\"a}mer}},
  \bibinfo{author}{\bibfnamefont{H.}~\bibnamefont{Mutka}},
  \bibinfo{author}{\bibfnamefont{A.}~\bibnamefont{Wildes}},
  \bibinfo{author}{\bibfnamefont{K.}~\bibnamefont{Habicht}}, \bibnamefont{and}
  \bibinfo{author}{\bibfnamefont{P.}~\bibnamefont{Vorderwisch}},
  \bibinfo{journal}{Nature} \textbf{\bibinfo{volume}{423}}, \bibinfo{pages}{62}
  (\bibinfo{year}{2003}).

\bibitem[{\citenamefont{Matsumoto et~al.}(2004)\citenamefont{Matsumoto,
  Normand, Rice, and Sigrist}}]{Matsumoto2004}
\bibinfo{author}{\bibfnamefont{M.}~\bibnamefont{Matsumoto}},
  \bibinfo{author}{\bibfnamefont{B.}~\bibnamefont{Normand}},
  \bibinfo{author}{\bibfnamefont{T.~M.} \bibnamefont{Rice}}, \bibnamefont{and}
  \bibinfo{author}{\bibfnamefont{M.}~\bibnamefont{Sigrist}},
  \bibinfo{journal}{Phys.\ Rev.\ {\bf B}} \textbf{\bibinfo{volume}{69}},
  \bibinfo{pages}{054423} (\bibinfo{year}{2004}).

\bibitem[{\citenamefont{Fleury and Loudon}(1968)}]{Fleury1968}
\bibinfo{author}{\bibfnamefont{P.~A.} \bibnamefont{Fleury}} \bibnamefont{and}
  \bibinfo{author}{\bibfnamefont{R.}~\bibnamefont{Loudon}},
  \bibinfo{journal}{Phys.\ Rev.} \textbf{\bibinfo{volume}{166}},
  \bibinfo{pages}{514} (\bibinfo{year}{1968}).

\bibitem[{\citenamefont{Glazkov et~al.}(2004)\citenamefont{Glazkov, Smirnov,
  Tanaka, and Oosawa}}]{Glazkov2004}
\bibinfo{author}{\bibfnamefont{V.~N.} \bibnamefont{Glazkov}},
  \bibinfo{author}{\bibfnamefont{A.~I.} \bibnamefont{Smirnov}},
  \bibinfo{author}{\bibfnamefont{H.}~\bibnamefont{Tanaka}}, \bibnamefont{and}
  \bibinfo{author}{\bibfnamefont{A.}~\bibnamefont{Oosawa}},
  \bibinfo{journal}{Phys.\ Rev.\ B} \textbf{\bibinfo{volume}{69}},
  \bibinfo{eid}{184410} (\bibinfo{year}{2004}).

\bibitem[{\citenamefont{Kusakabe et~al.}(2007)\citenamefont{Kusakabe, Kuroe,
  Oosawa, Sekine, Fujisawa, and Tanaka}}]{Kusakabe2007}
\bibinfo{author}{\bibfnamefont{K.}~\bibnamefont{Kusakabe}},
  \bibinfo{author}{\bibfnamefont{H.}~\bibnamefont{Kuroe}},
  \bibinfo{author}{\bibfnamefont{A.}~\bibnamefont{Oosawa}},
  \bibinfo{author}{\bibfnamefont{T.}~\bibnamefont{Sekine}},
  \bibinfo{author}{\bibfnamefont{M.}~\bibnamefont{Fujisawa}}, \bibnamefont{and}
  \bibinfo{author}{\bibfnamefont{H.}~\bibnamefont{Tanaka}},
  \bibinfo{journal}{J. Mag. Mag. Mater.} \textbf{\bibinfo{volume}{310}},
  \bibinfo{pages}{1365} (\bibinfo{year}{2007}).

\bibitem[{\citenamefont{Choi et~al.}(2003)\citenamefont{Choi, G\"{u}ntherodt,
  Oosawa, Tanaka, and Lemmens}}]{Choi2003}
\bibinfo{author}{\bibfnamefont{K.-Y.} \bibnamefont{Choi}},
  \bibinfo{author}{\bibfnamefont{G.}~\bibnamefont{G\"{u}ntherodt}},
  \bibinfo{author}{\bibfnamefont{A.}~\bibnamefont{Oosawa}},
  \bibinfo{author}{\bibfnamefont{H.}~\bibnamefont{Tanaka}}, \bibnamefont{and}
  \bibinfo{author}{\bibfnamefont{P.}~\bibnamefont{Lemmens}},
  \bibinfo{journal}{Phys.\ Rev.\ B} \textbf{\bibinfo{volume}{68}},
  \bibinfo{pages}{174412} (\bibinfo{year}{2003}).

\bibitem[{\citenamefont{Sekine et~al.}(1990)\citenamefont{Sekine, Jouanne,
  Julien, and Balkanski}}]{FePS3}
\bibinfo{author}{\bibfnamefont{T.}~\bibnamefont{Sekine}},
  \bibinfo{author}{\bibfnamefont{M.}~\bibnamefont{Jouanne}},
  \bibinfo{author}{\bibfnamefont{C.}~\bibnamefont{Julien}}, \bibnamefont{and}
  \bibinfo{author}{\bibfnamefont{M.}~\bibnamefont{Balkanski}},
  \bibinfo{journal}{Phys.\ Rev.\ B} \textbf{\bibinfo{volume}{42}},
  \bibinfo{pages}{8382} (\bibinfo{year}{1990}).

\bibitem[{\citenamefont{Kuroe et~al.}(1997)\citenamefont{Kuroe, Sasaki, Sekine,
  Koide, Sasago, Uchinokura, and Hase}}]{Kuroe1997}
\bibinfo{author}{\bibfnamefont{H.}~\bibnamefont{Kuroe}},
  \bibinfo{author}{\bibfnamefont{J.}~\bibnamefont{Sasaki}},
  \bibinfo{author}{\bibfnamefont{T.}~\bibnamefont{Sekine}},
  \bibinfo{author}{\bibfnamefont{N.}~\bibnamefont{Koide}},
  \bibinfo{author}{\bibfnamefont{Y.}~\bibnamefont{Sasago}},
  \bibinfo{author}{\bibfnamefont{K.}~\bibnamefont{Uchinokura}},
  \bibnamefont{and} \bibinfo{author}{\bibfnamefont{M.}~\bibnamefont{Hase}},
  \bibinfo{journal}{Phys.\ Rev.\ B} \textbf{\bibinfo{volume}{55}},
  \bibinfo{pages}{409} (\bibinfo{year}{1997}).

\bibitem[{\citenamefont{Halley}(1978)}]{Halley1978}
\bibinfo{author}{\bibfnamefont{J.~W.} \bibnamefont{Halley}},
  \bibinfo{journal}{Phys.\ Rev.\ Lett.} \textbf{\bibinfo{volume}{41}},
  \bibinfo{pages}{1605} (\bibinfo{year}{1978}).

\bibitem[{\citenamefont{Matsumoto et~al.}(2002)\citenamefont{Matsumoto,
  Normand, Rice, and Sigrist}}]{PhysRevLett.89.077203}
\bibinfo{author}{\bibfnamefont{M.}~\bibnamefont{Matsumoto}},
  \bibinfo{author}{\bibfnamefont{B.}~\bibnamefont{Normand}},
  \bibinfo{author}{\bibfnamefont{T.~M.} \bibnamefont{Rice}}, \bibnamefont{and}
  \bibinfo{author}{\bibfnamefont{M.}~\bibnamefont{Sigrist}},
  \bibinfo{journal}{Phys. Rev. Lett.} \textbf{\bibinfo{volume}{89}},
  \bibinfo{pages}{077203} (\bibinfo{year}{2002}).

\bibitem[{\citenamefont{Oosawa et~al.}(2002)\citenamefont{Oosawa, Kato, Tanaka,
  Kakurai, M\"uller, and Mikeska}}]{Oosawa2002}
\bibinfo{author}{\bibfnamefont{A.}~\bibnamefont{Oosawa}},
  \bibinfo{author}{\bibfnamefont{T.}~\bibnamefont{Kato}},
  \bibinfo{author}{\bibfnamefont{H.}~\bibnamefont{Tanaka}},
  \bibinfo{author}{\bibfnamefont{K.}~\bibnamefont{Kakurai}},
  \bibinfo{author}{\bibfnamefont{M.}~\bibnamefont{M\"uller}}, \bibnamefont{and}
  \bibinfo{author}{\bibfnamefont{H.-J.} \bibnamefont{Mikeska}},
  \bibinfo{journal}{Phys. Rev. B} \textbf{\bibinfo{volume}{65}},
  \bibinfo{pages}{094426} (\bibinfo{year}{2002}).

\bibitem[{\citenamefont{Takatsu et~al.}(1997)\citenamefont{Takatsu, Shiramura,
  and Tanaka}}]{Takatsu1997}
\bibinfo{author}{\bibfnamefont{K.}~\bibnamefont{Takatsu}},
  \bibinfo{author}{\bibfnamefont{W.}~\bibnamefont{Shiramura}},
  \bibnamefont{and} \bibinfo{author}{\bibfnamefont{H.}~\bibnamefont{Tanaka}},
  \bibinfo{journal}{J. Phys. Soc. Jpn.} \textbf{\bibinfo{volume}{66}},
  \bibinfo{pages}{1611} (\bibinfo{year}{1997}).

\bibitem[{\citenamefont{Weber and Ford}(1989)}]{Weber1989}
\bibinfo{author}{\bibfnamefont{W.~H.} \bibnamefont{Weber}} \bibnamefont{and}
  \bibinfo{author}{\bibfnamefont{G.~W.} \bibnamefont{Ford}},
  \bibinfo{journal}{Phys.\ Rev.\ B} \textbf{\bibinfo{volume}{40}},
  \bibinfo{pages}{6890} (\bibinfo{year}{1989}).

\bibitem[{\citenamefont{Knoll et~al.}(1990)\citenamefont{Knoll, Thomsen,
  Cardona, and Murugaraj}}]{PhysRevB.42.4842}
\bibinfo{author}{\bibfnamefont{P.}~\bibnamefont{Knoll}},
  \bibinfo{author}{\bibfnamefont{C.}~\bibnamefont{Thomsen}},
  \bibinfo{author}{\bibfnamefont{M.}~\bibnamefont{Cardona}}, \bibnamefont{and}
  \bibinfo{author}{\bibfnamefont{P.}~\bibnamefont{Murugaraj}},
  \bibinfo{journal}{Phys. Rev. B} \textbf{\bibinfo{volume}{42}},
  \bibinfo{pages}{4842} (\bibinfo{year}{1990}).

\bibitem[{\citenamefont{Sekine et~al.}(1998)\citenamefont{Sekine, Kuroe,
  Sasaki, Sasago, Koide, Uchinokura, and Hase}}]{Sekine1998}
\bibinfo{author}{\bibfnamefont{T.}~\bibnamefont{Sekine}},
  \bibinfo{author}{\bibfnamefont{H.}~\bibnamefont{Kuroe}},
  \bibinfo{author}{\bibfnamefont{J.}~\bibnamefont{Sasaki}},
  \bibinfo{author}{\bibfnamefont{Y.}~\bibnamefont{Sasago}},
  \bibinfo{author}{\bibfnamefont{N.}~\bibnamefont{Koide}},
  \bibinfo{author}{\bibfnamefont{K.}~\bibnamefont{Uchinokura}},
  \bibnamefont{and} \bibinfo{author}{\bibfnamefont{M.}~\bibnamefont{Hase}},
  \bibinfo{journal}{J.\ Phys.\ Soc.\ Jpn.} \textbf{\bibinfo{volume}{67}},
  \bibinfo{pages}{1440} (\bibinfo{year}{1998}).

\bibitem[{\citenamefont{Els et~al.}(1997)\citenamefont{Els, van Loosdrecht,
  Lemmens, Vonberg, G\"untherodt, Uhrig, Fujita, Akimitsu, Dhalenne, and
  Revcolevschi}}]{PhysRevLett.79.5138}
\bibinfo{author}{\bibfnamefont{G.}~\bibnamefont{Els}},
  \bibinfo{author}{\bibfnamefont{P.~H.~M.} \bibnamefont{van Loosdrecht}},
  \bibinfo{author}{\bibfnamefont{P.}~\bibnamefont{Lemmens}},
  \bibinfo{author}{\bibfnamefont{H.}~\bibnamefont{Vonberg}},
  \bibinfo{author}{\bibfnamefont{G.}~\bibnamefont{G\"untherodt}},
  \bibinfo{author}{\bibfnamefont{G.~S.} \bibnamefont{Uhrig}},
  \bibinfo{author}{\bibfnamefont{O.}~\bibnamefont{Fujita}},
  \bibinfo{author}{\bibfnamefont{J.}~\bibnamefont{Akimitsu}},
  \bibinfo{author}{\bibfnamefont{G.}~\bibnamefont{Dhalenne}}, \bibnamefont{and}
  \bibinfo{author}{\bibfnamefont{A.}~\bibnamefont{Revcolevschi}},
  \bibinfo{journal}{Phys. Rev. Lett.} \textbf{\bibinfo{volume}{79}},
  \bibinfo{pages}{5138} (\bibinfo{year}{1997}).

\bibitem[{\citenamefont{Choi et~al.}(2005)\citenamefont{Choi, Oosawa, Tanaka,
  and Lemmens}}]{Choi2005PRB}
\bibinfo{author}{\bibfnamefont{K.-Y.} \bibnamefont{Choi}},
  \bibinfo{author}{\bibfnamefont{A.}~\bibnamefont{Oosawa}},
  \bibinfo{author}{\bibfnamefont{H.}~\bibnamefont{Tanaka}}, \bibnamefont{and}
  \bibinfo{author}{\bibfnamefont{P.}~\bibnamefont{Lemmens}},
  \bibinfo{journal}{Phys.\ Rev.\ B} \textbf{\bibinfo{volume}{72}},
  \bibinfo{pages}{024451} (\bibinfo{year}{2005}).

\bibitem[{Gru()}]{Gruner1994}
\bibinfo{note}{G. Gr\"{u}ner, {\it Density Waves in Solids} (Addison-Wesley,
  Reading, MA, 1994), Chap. 6.}

\bibitem[{\citenamefont{Matsumoto et~al.}(2008)\citenamefont{Matsumoto, Kuroe,
  Oosawa, and Sekine}}]{Matsumoto2008}
\bibinfo{author}{\bibfnamefont{M.}~\bibnamefont{Matsumoto}},
  \bibinfo{author}{\bibfnamefont{H.}~\bibnamefont{Kuroe}},
  \bibinfo{author}{\bibfnamefont{A.}~\bibnamefont{Oosawa}}, \bibnamefont{and}
  \bibinfo{author}{\bibfnamefont{T.}~\bibnamefont{Sekine}},
  \bibinfo{journal}{J.\ Phys.\ Soc.\ Jpn.} \textbf{\bibinfo{volume}{77}},
  \bibinfo{pages}{033702} (\bibinfo{year}{2008}).

\bibitem[{Mat()}]{MatsumotoFuture}
\bibinfo{note}{M. Matsumoto {\it et al.}, unpublished.}

\bibitem[{\citenamefont{Goto et~al.}(2004)\citenamefont{Goto, Fujisawa, Ono,
  Tanaka, and Uwatoko}}]{Goto2004}
\bibinfo{author}{\bibfnamefont{K.}~\bibnamefont{Goto}},
  \bibinfo{author}{\bibfnamefont{M.}~\bibnamefont{Fujisawa}},
  \bibinfo{author}{\bibfnamefont{T.}~\bibnamefont{Ono}},
  \bibinfo{author}{\bibfnamefont{H.}~\bibnamefont{Tanaka}}, \bibnamefont{and}
  \bibinfo{author}{\bibfnamefont{Y.}~\bibnamefont{Uwatoko}},
  \bibinfo{journal}{J.\ Phys.\ Soc.\ Jpn.} \textbf{\bibinfo{volume}{73}},
  \bibinfo{pages}{3254} (\bibinfo{year}{2004}).

\bibitem[{\citenamefont{Oosawa et~al.}(2003)\citenamefont{Oosawa, Fujisawa,
  Osakabe, Kakurai, and Tanaka}}]{Oosawa2003}
\bibinfo{author}{\bibfnamefont{A.}~\bibnamefont{Oosawa}},
  \bibinfo{author}{\bibfnamefont{M.}~\bibnamefont{Fujisawa}},
  \bibinfo{author}{\bibfnamefont{T.}~\bibnamefont{Osakabe}},
  \bibinfo{author}{\bibfnamefont{K.}~\bibnamefont{Kakurai}}, \bibnamefont{and}
  \bibinfo{author}{\bibfnamefont{H.}~\bibnamefont{Tanaka}},
  \bibinfo{journal}{J.\ Phys.\ Soc.\ Jpn.} \textbf{\bibinfo{volume}{72}},
  \bibinfo{pages}{1026} (\bibinfo{year}{2003}).

\bibitem[{\citenamefont{Matsumoto and Koga}(2007)}]{Matsumoto2007}
\bibinfo{author}{\bibfnamefont{M.}~\bibnamefont{Matsumoto}} \bibnamefont{and}
  \bibinfo{author}{\bibfnamefont{M.}~\bibnamefont{Koga}}, \bibinfo{journal}{J.\
  Phys.\ Soc.\ Jpn.} \textbf{\bibinfo{volume}{76}}, \bibinfo{pages}{073709}
  (\bibinfo{year}{2007}).

\end{thebibliography}

\end{document}